\documentclass[aps, twocolumn, prb, floatfix, superscriptaddress, longbibliography, 10pt]{revtex4-2}
\usepackage{ragged2e}
\usepackage{ulem}
\usepackage{MnSymbol}
\usepackage{graphicx}
\usepackage{dcolumn}
\usepackage{bm}
\usepackage{mathtools}
\usepackage[export]{adjustbox}
\newcommand{\ope}{\operatorname}

\usepackage[colorlinks=true, linkcolor=black, citecolor=black, urlcolor=blue]{hyperref}

\begin{document}

\title{Mapping the non-equilibrium interacting Anderson Impurity Model\\ to an effective Gaussian theory}

\author{Emmanuel Bogacz}
\affiliation{School of Physics, University College Dublin, Belfield, Dublin 4, Ireland}
\affiliation{Centre for Quantum Engineering, Science, and Technology, University College Dublin, Dublin 4, Ireland}

\author{Graham Kells}
\affiliation{Department of Physics, Maynooth University, Maynooth, Co. Kildare, Ireland}
\affiliation{School of Theoretical Physics, Dublin Institute for Advanced Studies, 10 Burlington Road, Dublin 4, Ireland}

\author{Andrew K. Mitchell}
\email{Andrew.Mitchell@ucd.ie}
\affiliation{School of Physics, University College Dublin, Belfield, Dublin 4, Ireland}
\affiliation{Centre for Quantum Engineering, Science, and Technology, University College Dublin, Dublin 4, Ireland}

%%%%%%%%%%%%%%%%%%%%%%%%%%
%%%%%%%%%%%%%%%%%%%%%%%%%%

\begin{abstract}
Quantum impurity models with strong electron correlations, such as the paradigmatic Anderson Impurity Model (AIM), are central to our understanding of a range of physical phenomena including local moment formation, Coulomb blockade and Kondo screening. They describe magnetic atoms and molecules on surfaces, quantum dot circuits, and correlated materials through dynamical mean field theory. The physics of such systems in strongly non-equilibrium conditions is particularly complex and challenging to capture, whereas Gaussian models of free fermions can be easily solved. Here we show that the time-evolving dynamics of the AIM after a quench can be described by a completely non-interacting version of the model, at the expense of coupling to additional static auxiliary degrees of freedom. Starting from the full solution of the quenched AIM using ED and DMRG, we study the properties of this mapping using numerical optimization, 
and uncover intriguing structure in the auxiliary system. The method allows us to understand interacting non-equilibrium dynamics through the simpler lens of an effective non-interacting system of larger dimension.
\end{abstract}
\maketitle

%%%%%%%%%%%%%%%%%%%%%%%%%%
%%%%%%%%%%%%%%%%%%%%%%%%%%

\section{\label{sec:level1}Introduction}
The real-time dynamics of interacting many-body systems out of equilibrium is notoriously complex. Computational methods that work accurately and efficiently in equilibrium are often challenging or expensive to generalize in the non-equilibrium setting, due to the growth of temporal correlations and the absence of a simplifying thermal structure~\cite{Polkovnikov2011,Eisert2015,Aoki2014,schollwock2011density,Vidal2003,Calabrese2005}. 

Quantum impurity models~\cite{Hewson1993} describe an important but simpler class of strongly-correlated physics, in which an interacting system (the `impurity' degrees of freedom) is coupled to an infinite non-interacting bath. They may be regarded as non-Markovian open quantum systems in strong coupling, where strong multipartite system-bath entanglement develops at low temperatures due to the Kondo effect~\cite{Sorensen2007,Lee2015,Shim2018}, requiring explicit treatment of the system-bath composite. Such models describe actual magnetic impurities in metals~\cite{Hewson1993,costi2009kondo,derry2015quasiparticle} or superconductors~\cite{Satori1992,Zitko2016,Huang2023} and molecules on surfaces~\cite{Chiappe2006,DiasDaSilva2009,li2025negative}; as well as quantum circuits involving quantum dots~\cite{goldhaber1998kondo,cronenwett1998tunable,jeong2001kondo,mitchell2012universal}, Coulomb boxes~\cite{lebanon2003coulomb,potok2007observation} or hybrid metal-semiconductor sites~\cite{iftikhar2018tunable,pouse2023quantum,sriram2025hybrid} and single-molecule transistors~\cite{liang2002kondo,perrin2015single,mitchell2017kondo}. They are also the effective local models describing correlated materials in the context of dynamical mean field theory (DMFT)~\cite{georges1996dynamical,Aoki2014}.

In equilibrium, numerically-exact solutions can be obtained using sophisticated computational techniques such as Wilson's Numerical Renormalization Group (NRG)~\cite{wilson1975renormalization,krishna1980renormalization,weichselbaum2007sum,mitchell2014generalized}, continuous-time quantum Monte Carlo~\cite{gull2011continuous,werner2006continuous}, and tensor network methods such as the Density Matrix Renormalization Group (DMRG)~\cite{white1992density,schollwock2011density}. Out of equilibrium, the development of time-dependent NRG (TDNRG)~\cite{nghiem2017time, nghiem2018time}, the Time Dependent Variational Principle (TDVP) within 
DMRG~\cite{haegman2011time}, and more recent methods involving process tensors~\cite{link2024open} or 
neural quantum states~\cite{Lange_2024},
has enabled the first exact explorations of interacting quantum impurity systems subject to external time-dependent driving. Such studies have implications, for example, in quantum transport~\cite{schwarz2016lindblad} and non-equilibrium quantum thermodynamics~\cite{campbell2026roadmap}. However, such numerical studies are highly computationally demanding, and a clean physical understanding of the underlying dynamical behavior may not be apparent.

The simplest example is the single-impurity, single-channel Anderson impurity model (AIM)~\cite{Hewson1993},
\begin{eqnarray}\label{eq:AIM}
\begin{split}
H_{\mathrm{AIM}}
=&
\sum_{\sigma} \epsilon_d(t) \, d_{\sigma}^{\dagger} d_{\sigma} +
U d_{\uparrow}^{\dagger} d_{\uparrow}d_{\downarrow}^{\dagger} d_{\downarrow}+\sum_{k\sigma} \epsilon_k \, c_{k\sigma}^{\dagger} c_{k\sigma} \\
&+ \sum_{k\sigma}
\left(
V_k c_{k\sigma}^{\dagger} d_{\sigma}
+
V_k^{*} d_{\sigma}^{\dagger} c_{k\sigma}
\right) \;,
\end{split}
\end{eqnarray}
where $d_{\sigma}^{(\dagger)}$ are annihilation (creation) operators for impurity electrons and $c_{k\sigma}^{(\dagger)}$ are annihilation (creation) operators for bath electrons of momentum $k$; with $\sigma=\uparrow/\downarrow$ the spin. 
Time-dependence is introduced through the impurity potential $\epsilon_d(t)$, and here we specialize further to the quench $\epsilon_d(t)=\epsilon_d^i+\delta \theta(t)$, where the potential is changed suddenly at $t=0$ from $\epsilon_d^i$ to $\epsilon_d^f=\epsilon_d^i+\delta$.

The non-interacting $U\to 0$ limit of the AIM is referred to as the resonant level model (RLM), and can be solved exactly. In this case, an impurity potential quench preserves the quadratic (Gaussian) structure of the problem, and Wick’s theorem can be applied. The dynamics can then be obtained exactly from the single-particle evolution. In the full AIM, the interaction $U>0$ precludes such a reduction to independent single-particle dynamics and the interaction self-energy must be accounted for.

In the interacting but equilibrium limit $\delta\to 0$ (no quench), it was shown in Refs.~\cite{sen2020mott,PW_aux}
that a fully non-interacting model can be systematically  constructed that \textit{exactly}  reproduces the same system dynamics (for example, the impurity Green's function). This comes at the cost of embedding the original AIM with $U=0$ into a higher-dimensional Fock space, by coupling to additional auxiliary fermionic degrees of freedom. The advantage is that the effective Gaussian theory is trivial to analyze, with complexity growing linearly in system size, rather than exponentially. The mapping is deterministic but requires prior knowledge of the interaction self-energy for the original AIM; its interest lies in the novel perspective on the complex dynamics of the interacting system through the lens of simpler single-particle physics. For example in the context of DMFT, one can view the Mott transition in the Hubbard model as a \textit{topological} transition in this auxiliary space~\cite{sen2020mott}.

%%%%%%%%%%%%%%%%%%
%%%%%%%%%%%%%%%%%%

\begin{figure}[t]
    \centering
    \includegraphics[width=0.95\linewidth]{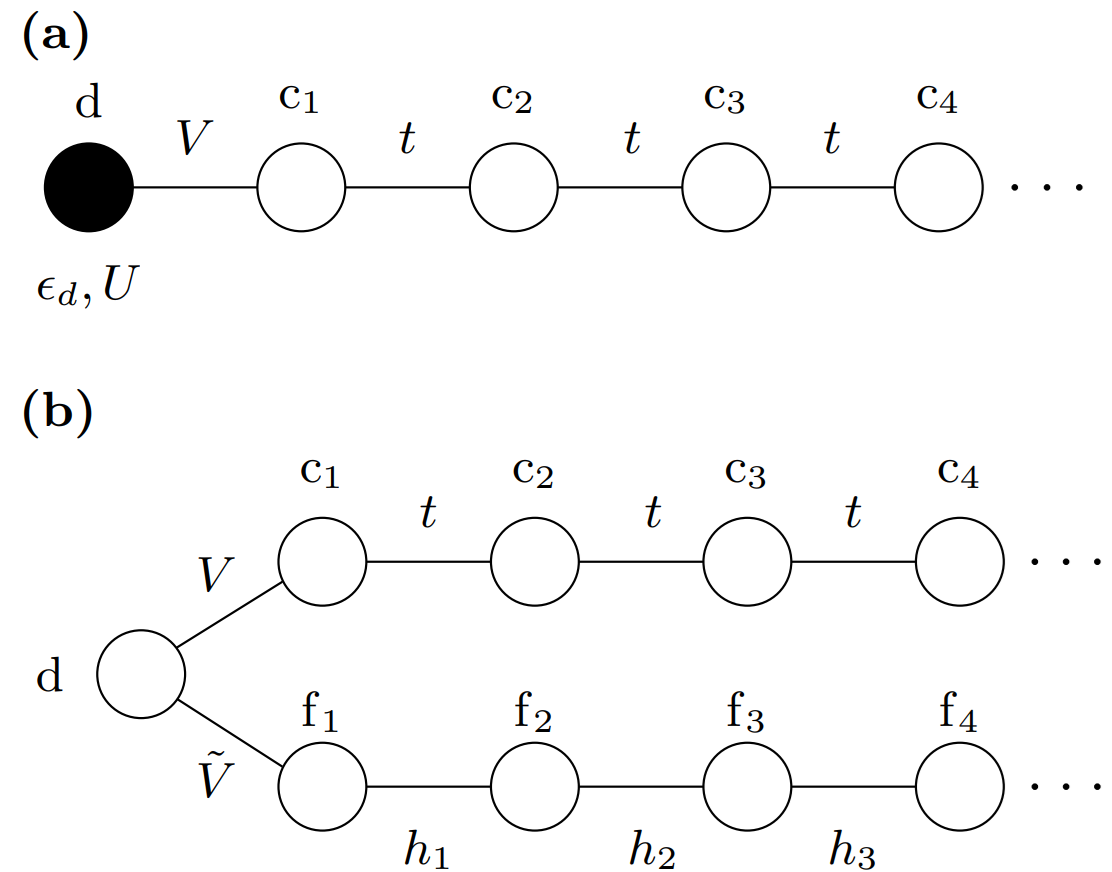}
\caption{Schematic representations of the bare and effective models in chain form. (a) Interacting AIM; and (b) Non-interacting effective model in which $U=0$ but the impurity couples to an additional auxiliary chain of fermions $f_n$. }
\label{cartoons}
\end{figure}

%%%%%%%%%%%%%%%%%%
%%%%%%%%%%%%%%%%%%

Out of equilibrium, it is not obvious whether such a mapping can be constructed. For the quenched AIM, can one formulate an effective Gaussian theory where the non-equilibrium real-time dynamics of system observables are reproduced by simply quenching the auxiliary system? If so, what insights can be gained from the emergent structure of the auxiliary system?

In this paper, we demonstrate explicitly that in fact equivalent non-interacting impurity models \textit{can be found} that accurately reproduce the quench dynamics of the interacting AIM. We determine the mapping numerically using numerical optimization techniques, starting from DMRG solutions of the original problem. We consider the required complexity and dimension of the auxiliary system to achieve accurate evolution at a given timescale, and study the properties of this auxiliary system. We show that results are systematically improvable. We do not claim that the mapping provides a shortcut to solving the non-equilibrium dynamics of the original interacting system. However, optimizing the parameters of our physically-motivated effective model may provide an efficient and more interpretable route to machine learning the time-dependence of observables in complex systems.

The paper is organized as follows. In Sec.~\ref{sec:equilib} we present the bare and effective models in chain form, and describe the mapping between them in equilibrium. In Sec.~\ref{sec:quench} we consider the time-dependence of the impurity occupation after a quench, and the auxiliary chain mapping using numerical optimization methods. Properties of the auxiliary chain are studied. A discussion and outlook is given in Sec.~\ref{sec:conc}. Technical details are provided in the appendix.

%%%%%%%%%%%%%%%%%%
%%%%%%%%%%%%%%%%%%

\section{Equilibrium mapping}\label{sec:equilib}
We consider first the mapping of an interacting AIM to an effective non-interacting auxiliary chain representation (ACR) in equilibrium at zero temperature. 

\subsection{Bare model}
We take a chain form for the original AIM, with a finite number of $N$ bath sites. This is a particular case of Eq.~\ref{eq:AIM}, written explicitly as:
\begin{eqnarray}\label{eq:AIMchain}
\begin{split}
H_{\mathrm{AIM}}
=&
\sum_{\sigma} \epsilon_d^{\phantom{\dagger}} \, d_{\sigma}^{\dagger} d_{\sigma}^{\phantom{\dagger}}    
+
U d_{\uparrow}^{\dagger} d_{\uparrow}d_{\downarrow}^{\dagger} d_{\downarrow} + V \sum_{\sigma} \left( d_{\sigma}^{\dagger}c_{1\sigma}^{\phantom{\dagger}}+c_{1\sigma}^{\dagger} d_{\sigma}^{\phantom{\dagger}} \right)  \\ &
+t\sum_{n=1}^{N-1}\sum_{\sigma} \left( c_{n\sigma}^{\dagger}c_{n+1,\sigma}^{\phantom{\dagger}} +c_{n+1,\sigma}^{\dagger}c_{n\sigma}^{\phantom{\dagger}} \right
) \;.
\end{split}
\end{eqnarray}
The setup is illustrated in Fig.~\ref{cartoons}(a). This model can be fully solved up to around $N=10$ using standard exact diagonalization (ED)~\cite{weisse2008exact}, and ground state properties can be straightforwardly obtained up to $N\sim 200$ or more using DMRG~\cite{schollwock2011density}. The thermodynamic limit $N\to \infty$ can be treated with NRG~\cite{bulla2008numerical}. We use these methods to examine the different finite-size behaviors of the mapping.

%%%

\subsubsection{Representations of the Interacting Green's function}
The impurity retarded Green's function is defined as $G^{\rm AIM}_{dd;\sigma}(t)=-i\theta(t)\langle \{d_{\sigma}(t),d_{\sigma}^{\dagger}(0) \}\rangle$ where $\langle\dots\rangle$ is the thermal expectation value in the Gibbs state. In the (complex) frequency domain $z=\omega+i\eta$ with $\eta>0$ we have $G^{\rm AIM}_{dd;\sigma}(z)\!\equiv\! \langle \langle d_{\sigma};d_{\sigma}^{\dagger}\rangle\rangle\!=\!\int dt \:e^{izt} G^{\rm AIM}_{dd;\sigma}(t)$.
Its spectral function is  $A(\omega)=-\tfrac{1}{\pi}{\rm Im}\:G^{\rm AIM}_{dd;\sigma}(\omega+i0^+)$ and the impurity occupation for spin-$\sigma$ is $\langle n_{d\sigma}\rangle=\int d\omega \: A(\omega) f(\omega)$ with $f(\omega)$ the Fermi function and $n_{d\sigma}=d^{\dagger}_{\sigma}d_{\sigma}$. Given spin symmetry we have $\langle n_{d\uparrow}\rangle=\langle n_{d\downarrow}\rangle$ and we denote the total impurity occupation $\langle n_{d}\rangle=\sum_{\sigma}\langle n_{d\sigma}\rangle$.

In the $U=0$ non-interacting limit, the Green's function of the resulting RLM takes the form,
\begin{equation}\label{eq:rlm}
G^{\rm RLM}_{dd;\sigma}(z) = \frac{1}{z-\epsilon_d-\Delta(z)} \;,
\end{equation}
where the hybridization function $\Delta(z)$ takes an exact continued-fraction form obtained from equations of motion methods,
\begin{equation}
\Delta(z) =
\frac{V^2}{
z-
\dfrac{t^2}{
z-
\dfrac{t^2}{
z-
\dfrac{t^2}{z-\cdots}
}
}
} \;,
\label{eq:hyb}
\end{equation}
which has $N$ levels for a system with $N$ bath sites.

For $U>0$ the interaction self-energy $\Sigma_{dd;\sigma}^{\rm AIM}(z)$ must be accounted for, resulting in the Dyson form,
\begin{equation}\label{eq:dyson}
G^{\rm AIM}_{dd;\sigma}(z) = \frac{1}{z-\epsilon_d-\Delta(z)-\Sigma_{dd;\sigma}^{\rm AIM}(z)} \;.
\end{equation}
The self-energy can also be expanded as a continued fraction of $M\gg N$ levels, as shown in Refs.~\cite{sen2020mott,PW_aux},
\begin{equation}
\Sigma_{dd;\sigma}^{\rm AIM}(z) =\Sigma_{\rm HF}+
\frac{\tilde{V}^2}{
z-e_1-
\dfrac{h_1^2}{
z-e_2-
\dfrac{h_2^2}{
z-e_3-
\dfrac{h_3^2}{z-\cdots}
}
}
} \;,
\label{eq:SE_cfe}
\end{equation}
with the static Hartree-Fock piece $\Sigma_{\rm HF}=U\langle n_{d\bar{\sigma}}\rangle$ and $\tilde{V}=U\sqrt{\langle n_{d\bar{\sigma}}\rangle(1-\langle n_{d\bar{\sigma}}\rangle)}$. The remaining coefficients $\{h_n\}$ and $\{e_n\}$ are nontrivial, and encode information about the higher moments of the self-energy. This structure will be exploited in the effective auxiliary-chain mapping described in the next subsection. 

Related continued-fraction forms for the Green's function are obtained directly from the Lehmann representation~\cite{bruus2004many} expanded in a Krylov basis~\cite{Gagliano1987}. At zero temperature, we decompose the Green's function into particle addition and removal processes from the many-body $m$-electron ground state $|\Psi_0^m\rangle$, via: $G^{\rm AIM}_{dd;\sigma}(z)=G^+(z)+G^-(z)$. Their Lehmann representations read,
\begin{equation}\label{eq:Gpm}
    G^{\pm}(z)=\mu_0^{\pm}\sum_j\frac{\left|\left\langle \Psi^{m\pm 1}_j|q_0^{\pm}\right \rangle\right |^2}{z\pm (E_0^m-E_j^{m\pm 1}) }
\end{equation}
where $H_{\rm AIM}|\Psi^n_j\rangle=E^n_j|\Psi^n_j\rangle$ and $|q_0^\pm\rangle$ are normalized states obtained by adding or removing an electron from the ground state, defined by  $|q_0^+\rangle=d_{\sigma}^{\dagger}|\Psi^m_0\rangle/\sqrt{\mu_0^+}$ and $|q_0^-\rangle=d_{\sigma}|\Psi^m_0\rangle/\sqrt{\mu_0^-}$ with the  normalizations $\mu_0^{+}=\langle \Psi_0^m|d_{\sigma} d^{\dagger}_{\sigma}|\Psi_0^m\rangle=1-\langle n_{d\sigma}\rangle$ and $\mu_0^{-}=\langle \Psi_0^m|d^{\dagger}_{\sigma} d_{\sigma}|\Psi_0^m\rangle=\langle n_{d\sigma}\rangle$. This construction gives a sum of simple poles. Taken together the impurity spectral function is,
\begin{eqnarray}\label{eq:A}
    A(\omega)=\sum_i w_i\delta(\omega-\xi_i)
\end{eqnarray}
with the pole weights $w_i$ and positions $\xi_i$ related to the matrix elements and excitations energies in Eq.~\ref{eq:Gpm}.

Alternatively we can expand the resolvent in the Krylov basis to obtain a continued fraction form for each of $G^+(z)$ and $G^-(z)$. Starting instead from,
\begin{equation}
G^\pm(z) =
\mu_0^\pm
\langle q_0^\pm|
\frac{1}{z-K_\pm}
|q_0^\pm\rangle .
\label{eq:Gpm_res}
\end{equation}
with $K_\pm=\pm(P_{m\pm 1}H_{\rm AIM} P_{m\pm 1} - E_0^m)$ an effective projected Hamiltonian and $P_n=\sum_j |\Psi^n_j\rangle\langle \Psi^n_j|$, we define the Krylov subspaces generated by \(K_\pm\) and \(|q_0^\pm\rangle\) as,
\begin{equation}
\mathcal{K}_{D_\pm}^\pm =
\mathrm{span}
\left\{
|q_0^\pm\rangle,
K_\pm |q_0^\pm\rangle,
K_\pm^2 |q_0^\pm\rangle,
\ldots,
K_\pm^{D_\pm-1}|q_0^\pm\rangle
\right\} \;,
\end{equation}
where $D_\pm$ is the dimension of the $m\pm 1$ subspace. 
An orthonormal basis 
\(\{|q_n^\pm\rangle\}\) is then constructed using Lanczos tridiagonalization,
\begin{equation}
K_\pm |q_n^\pm\rangle
=
b_{n+1}^\pm |q_{n+1}^\pm\rangle
+
a_n^\pm |q_n^\pm\rangle
+
b_n^\pm |q_{n-1}^\pm\rangle 
\end{equation}
with $b_0^\pm=0$ and,
\begin{eqnarray}
a_n^\pm &=& \langle q_n^\pm|K_\pm|q_n^\pm\rangle \;,\\
b_{n+1}^\pm & =&\left\|
K_\pm |q_n^\pm\rangle
-
a_n^\pm |q_n^\pm\rangle
-
b_n^\pm |q_{n-1}^\pm\rangle
\right\| \;.
\end{eqnarray}
In this basis, \(K_\pm\) is represented by the tridiagonal  matrix
\begin{equation}
T^\pm
=
\begin{pmatrix}
a_0^\pm & b_1^\pm & 0 & \cdots \\
b_1^\pm & a_1^\pm & b_2^\pm & \cdots \\
0 & b_2^\pm & a_2^\pm & \ddots \\
\vdots & \vdots & \ddots & \ddots
\end{pmatrix}.
\end{equation}
The projected resolvent is therefore,
\begin{equation}
G^\pm(z) = \mu_0^\pm \left[z I-T^\pm\right]^{-1}_{11}
\end{equation}
which yields the continued-fraction form,
\begin{equation}
G^\pm(z) =
\frac{\mu_0^\pm}{
z-a_0^\pm
-
\dfrac{(b_1^\pm)^2}{
z-a_1^\pm
-
\dfrac{(b_2^\pm)^2}{
z-a_2^\pm
-
\dfrac{(b_3^\pm)^2}{z-a_3^\pm-\cdots}
}
}
}.
\label{eq:Gpm_continued_fraction}
\end{equation}
As such the physical impurity Green's function $G^{\rm AIM}_{dd;\sigma}(z)=G^+(z)+G^-(z)$ can be written as the sum of two continued fractions.

Less conventionally, but equally valid, one can formulate the above procedure in such a way that a single continued fraction is obtained directly. 
To do this we define the composite operator,
\begin{equation}
K_{\rm tot} = K_+ \oplus K_- \;,
\end{equation}
and corresponding composite initial vector,
\begin{equation}
|F_0\rangle = d_\sigma^\dagger|\Psi_0^m\rangle
\oplus
d_\sigma|\Psi_0^m\rangle \;,
\end{equation}
which is normalized, $\langle F_0|F_0\rangle
\!= \!\langle \{d_\sigma,d_\sigma^\dagger\}\rangle\! =\! 1$.
The full Green's function may therefore be written as the single composite resolvent,
\begin{equation}
G^{\rm AIM}_{dd;\sigma}(z)=
\langle F_0|
\frac{1}{z-K_{\rm tot}}
|F_0\rangle \;.
\label{eq:composite_resolvent}
\end{equation}
Lanczos tridiagonalization of \(K_{\rm tot}\), starting from the initial state vector \(|F_0\rangle\), then 
generates a single continued-fraction for the impurity Green's function, $G^{\rm AIM}_{dd;\sigma}(z)$. 

Indeed, the Green's function for any site can be expanded in this way using the same $K_{\rm tot}$, simply by choosing a different starting state vector. In particular, the Green's function at the \textit{other} end of the chain, $G^{\rm AIM}_{NN;\sigma}(z)=\langle\langle c^{\phantom{\dagger}}_{N\sigma} ; c_{N\sigma}^{\dagger}\rangle\rangle$ can be obtained by setting,
\begin{equation}
|\mathcal{F}_0\rangle = c_{N\sigma}^\dagger|\Psi_0^m\rangle
\oplus
c_{N\sigma}|\Psi_0^m\rangle \;,
\end{equation}
and then expanding 
\begin{equation}
G^{\rm AIM}_{NN;\sigma}(z)=
\langle \mathcal{F}_0|
\frac{1}{z-K_{\rm tot}}
|\mathcal{F}_0\rangle \;.
\end{equation}
This yields a single continued fraction,
\begin{equation}
G^{\rm AIM}_{NN;\sigma}(z) =
\frac{1}{
z-A_0
-
\dfrac{B_1^2}{
z-A_1
-
\dfrac{B_2^2}{
z-A_2
-
\dfrac{B_3^2}{z-A_3-\cdots}
}
}
}.
\label{eq:end_chain_cfe}
\end{equation}

Running this algorithm to obtain $G^{\rm AIM}_{NN;\sigma}(z)$ for our finite chain AIM Hamiltonian Eq.~\ref{eq:AIMchain}, we find that the first $N$ coefficients $A_n=0$ are exactly zero, and the first $N-1$ coefficients $B_n=t$ are given exactly by the bare lead hopping, whereas $B_N=V$. This is the (reversed) expansion of the bare hybridization, Eq.~\ref{eq:hyb}. The continued fraction then continues with nontrivial coefficients. These coefficients are found to be precisely those of the self-energy expansion, Eq.~\ref{eq:SE_cfe}. This curious structure hints at a specific geometry for an effective non-interacting model, as discussed in the next section. 

%%%%%%%%%%%%%%%%%%%%%%%
%%%%%%%%%%%%%%%%%%%%%%%

\subsection{Effective model with auxiliary chain}
Consider a fully non-interacting (quadratic) tight-binding chain model with the geometry depicted in Fig.~\ref{cartoons}(b). A site $d_{\sigma}$ is singled out as the impurity, but with $U=0$. It is coupled to the physical non-interacting bath of $N$ sites $c_{n\sigma}$ of the original AIM via the hopping $V$, and also to an `auxiliary chain' of $M$ additional free fermionic sites $f_{n\sigma}$ via the coupling $\tilde{V}$. This will be our effective model for the AIM. Its Hamiltonian reads,
\begin{eqnarray}\label{eq:Heff}
\begin{split}
H_{\mathrm{eff}}
=&
\sum_{\sigma} \tilde{\epsilon}_d^{\phantom{\dagger}} \, d_{\sigma}^{\dagger} d_{\sigma}^{\phantom{\dagger}}  + V \sum_{\sigma} \left( d_{\sigma}^{\dagger}c_{1\sigma}^{\phantom{\dagger}}+c_{1\sigma}^{\dagger} d_{\sigma}^{\phantom{\dagger}} \right)  \\ &
+t\sum_{n=1}^{N-1}\sum_{\sigma} \left( c_{n\sigma}^{\dagger}c_{n+1,\sigma}^{\phantom{\dagger}} +c_{n+1,\sigma}^{\dagger}c_{n\sigma}^{\phantom{\dagger}} \right
) \\
& + \tilde{V} \sum_{\sigma} \left( d_{\sigma}^{\dagger}f_{1\sigma}^{\phantom{\dagger}}+f_{1\sigma}^{\dagger} d_{\sigma}^{\phantom{\dagger}} \right) + \sum_{m=1}^{M}\sum_{\sigma} e_n f_{n\sigma}^{\dagger}f_{n\sigma}^{\phantom{\dagger}}  \\
&
+\sum_{m=1}^{M-1}\sum_{\sigma} h_n\left( f_{n\sigma}^{\dagger}f_{n+1,\sigma}^{\phantom{\dagger}} +f_{n+1,\sigma}^{\dagger}f_{n\sigma}^{\phantom{\dagger}} \right
) \;.
\end{split}
\end{eqnarray}
The first two lines describe the $U=0$ limit of the original AIM (with renormalized potential $\epsilon_d\to \tilde{\epsilon}_d$), whereas the last two lines describe coupling to the auxiliary chain, parametrized by onsite auxiliary potentials $e_n$ and hoppings $h_n$. Since the model is diagonal and symmetric in the spin label $\sigma$ one can in practice drop the spin label and consider spinless fermions.

The impurity Green's function of this system is readily obtained from equations of motion~\cite{Zubarev1960}, yielding a double continued-fraction form,
\begin{eqnarray}\label{eq:Heff_GF}
G^{\:\rm eff}_{dd;\sigma}(z) = \qquad\qquad\qquad\qquad\qquad\qquad\qquad\qquad\qquad \\
\frac{1}{
z-\tilde{\epsilon}_d
- \dfrac{V^2}{z-\dfrac{t^2}{z-\dfrac{t^2}{z-\cdots}}} - \dfrac{ \tilde{V}^2}{z-e_1-\dfrac{h_1^2}{z-e_2-\dfrac{h_2^2}{z-e_3-\cdots}}}
}\nonumber
\end{eqnarray}

We note that this is precisely the structure obtained for Eq.~\ref{eq:dyson} when the continued-fraction expansions for $\Delta(z)$ and $\Sigma^{\rm AIM}_{dd;\sigma}(z)$ are inserted from Eqs.~\ref{eq:hyb} and \ref{eq:SE_cfe}. The mapping is therefore achieved by setting $\tilde{\epsilon}_d=\epsilon_d+\tfrac{1}{2}U\langle n_d\rangle$ and $\tilde{V}=\tfrac{1}{2}U\sqrt{\langle n_d\rangle(2-\langle n_d\rangle)}$ as well as $\{e_n\}$ and $\{h_n\}$ obtained from Eq.~\ref{eq:SE_cfe}. Defining the non-interacting model Eq.~\ref{eq:Heff} with these parameters is guaranteed~\cite{sen2020mott,PW_aux} to exactly reproduce the full dynamics of the interacting AIM impurity Green's function, $G^{\:\rm eff}_{dd;\sigma}(z)=G^{\rm AIM}_{dd;\sigma}(z)$. 
Indeed, all Green's functions on the physical sites $d_{\sigma}$ and $c_{n\sigma}$, such as $\langle\langle d_{\sigma} ; c_{n\sigma}^{\dagger}\rangle \rangle$ or $\langle\langle c_{i\sigma} ; c_{j\sigma}^{\dagger}\rangle \rangle$, will exactly match in the interacting AIM and non-interacting effective model. This is because the equations of motion relating any physical Green's function to the impurity Green's function $G^{\:\rm eff}_{dd;\sigma}(z)$ are identical to the expressions for the the original AIM, due to the chain geometry. As an immediate consequence, the physical site occupations and single-particle correlation matrix also exactly match. By contrast, correlation functions involving the fictitious auxiliary sites, such as $\langle\langle d_{\sigma} ; f_{n\sigma}^{\dagger}\rangle \rangle$ or $\langle\langle f_{i\sigma} ; f_{j\sigma}^{\dagger}\rangle \rangle$ have no physical interpretation in the AIM.

What, then, is the physical meaning of the auxiliary degrees of freedom? Can the parameters of the effective model be found without computing the self-energy? 

The effective model Eq.~\ref{eq:Heff} consists of a single 1d chain, with the impurity a distance $N$ bath sites from the end. The double continued-fraction form of $G^{\:\rm eff}_{dd;\sigma}(z)$ in Eq.~\ref{eq:Heff_GF} owes to the fact that the chain extends away from the impurity site in both directions. If instead we compute $G^{\:\rm eff}_{NN;\sigma}(z)=\langle\langle c_{N\sigma} ; c_{N\sigma}^{\dagger}\rangle \rangle$ at the end of the chain, we obtain a single continued fraction. This is exactly the result obtained from the Lanczos/Krylov expansion of the \textit{interacting} Green's function of the AIM via the Lehmann representation, presented in the previous section, see Eq.~\ref{eq:end_chain_cfe}. Our effective model parameters can therefore be read off directly from the expansion of $G^{\rm AIM}_{NN;\sigma}(z)$. The single-particle auxiliary states of our effective model $|\phi_m\rangle=f_{m}^{\dagger}|\rm vac\rangle$ can therefore be understood as the Krylov basis states $|q_{N+m+2}\rangle$ of the AIM. 

The full number of auxiliary chain sites $M$ is the dimension of the AIM Krylov subspace, minus $N+1$ for the number of physical impurity and bath sites. This is (at most) the combined  dimension $D_++D_-$ of the many-body $m\pm 1$ electron sectors of the AIM. However, in practice one can often truncate the auxiliary chain at a smaller value of $M$ to get an approximation of $G^{\rm AIM}_{NN;\sigma}(z)$ via $H_{\rm eff}$. This is systematically improvable by adding more auxiliary sites, and convergence of the spectral functions can be checked \textit{post hoc}. 
Because additional auxiliary sites are rather distant from the impurity, these have progressively less impact on the target impurity spectral function, and so convergence is rather rapid; the chain (Lanczos) form naturally organizes the information hierarchically. 
A useful rule-of-thumb is that the total number of sites in the effective system ($N+M+1$) is the number of highest-weight poles in the resulting impurity spectral function. We revisit the question of the number of auxiliary sites needed for quenched systems later.

%%%%%%%%%%

\subsection{Numerical demonstration in equilibrium}
In equilibrium, our auxiliary chain mapping is well-defined and controlled, as explained above. Before moving on to non-equilibrium quench dynamics within the ACR, we first demonstrate the accuracy and expressivity of the mapping on the level of the impurity spectral function for equilibrium systems.

\begin{figure}
    \centering
    \includegraphics[width=0.98\linewidth]{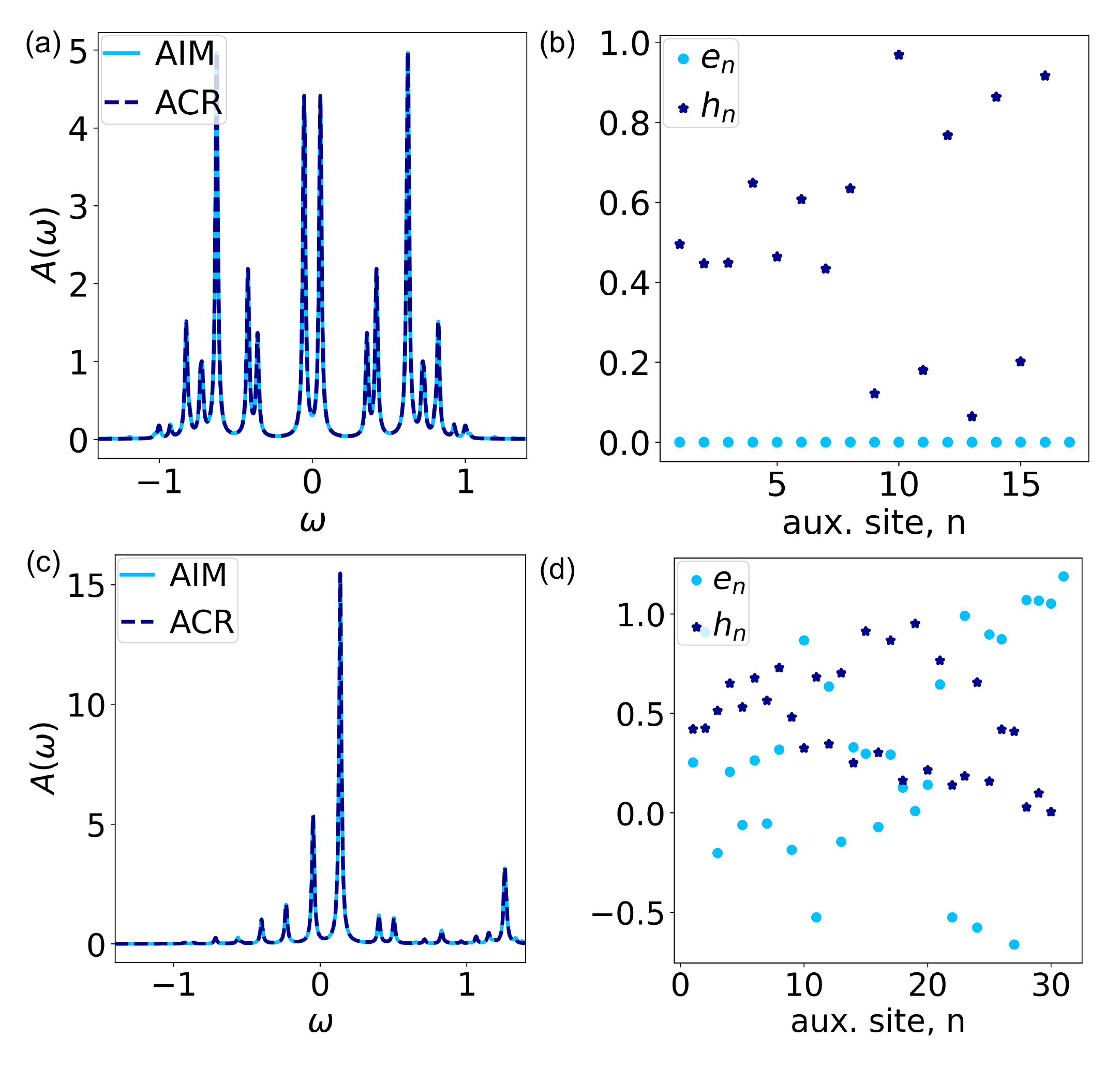}
    \caption{Auxiliary chain mapping in equilibrium for the AIM with $N=7$ bath sites. Panels (a,c) show the impurity spectral function $A(\omega)$ for the interacting AIM computed exactly with ED (light blue lines), and for the effective non-interacting ACR (dark blue lines). Panels (b,d) show the auxiliary chain parameters. Here $U=1$, $t=0.5$, $V=0.2$ and $\epsilon_d=-U/2$ (half filling) in (a,b) but $\epsilon_d=0$ (mixed valence) in (c,d).}
    \label{fig:eq_ed}
\end{figure}

\begin{figure}
    \centering
    \includegraphics[width=1\linewidth]{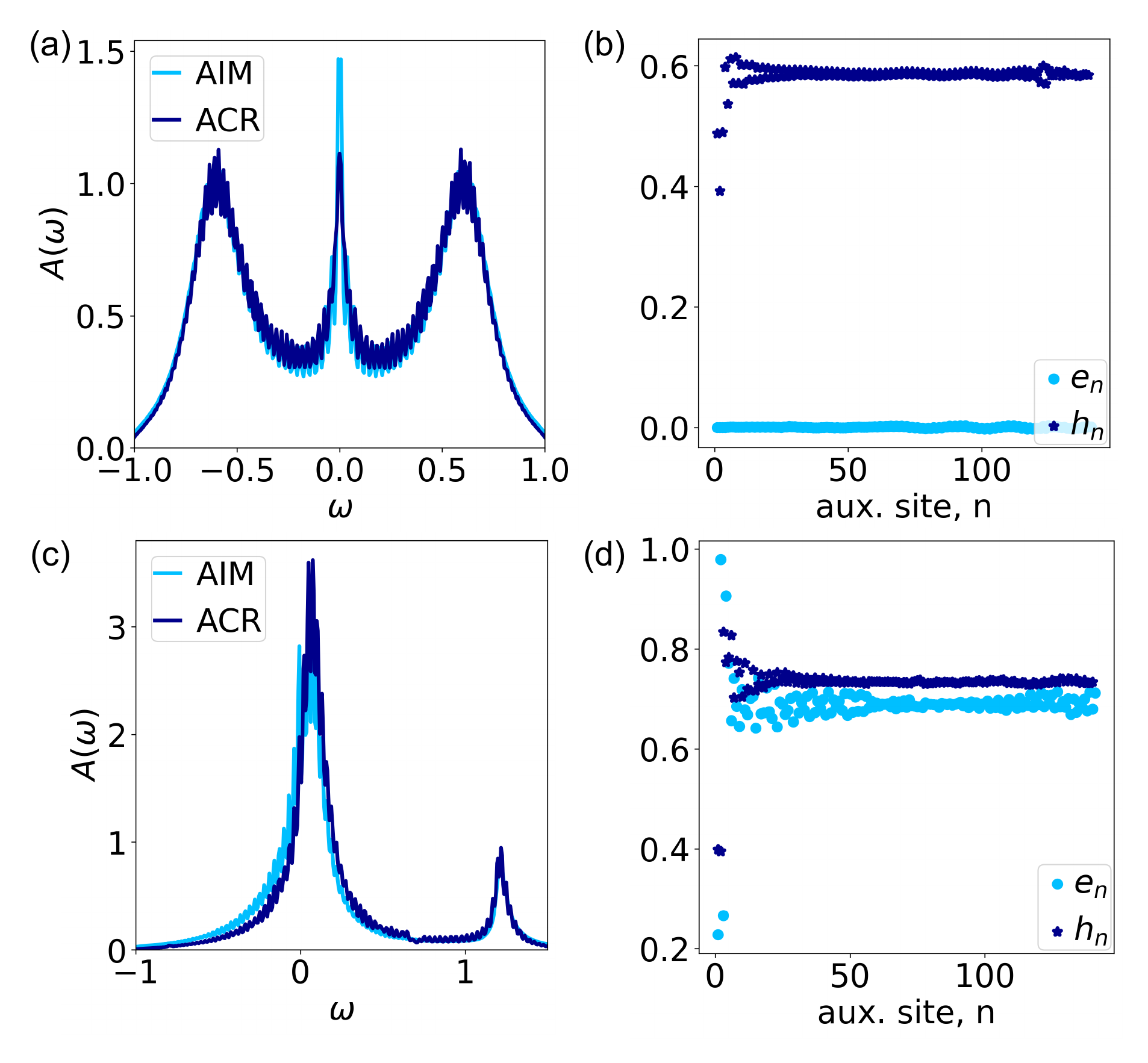}
    \caption{Same as Fig.~\ref{fig:eq_ed} but for $N=101$ bath sites, with the AIM calculation now performed using TDVP-MPS methods.}
    \label{fig:eq_mps}
\end{figure}

Fig.~\ref{fig:eq_ed} shows results for the equilibrium mapping from the interacting AIM to the non-interacting ACR. Here we consider an impurity coupled to $N=7$ physical bath sites, and solve the bare model exactly in the strongly-correlated regime using ED. We take representative model parameters $U=1$, $t=0.5$, $V=0.2$. The choice $t=0.5$ means a conduction electron bandwidth $D=1$ in the thermodynamic limit $N\to \infty$, and we use this as our unit of energy. Top panels (a,b) are for the half-filling case $\epsilon_d=-U/2$ whereas bottom panels (c,d) are for the mixed-valence case $\epsilon_d=0$. We keep these same set of parameters throughout the paper to afford a direct comparison of results (but vary $N$). Left panels (a,c) show the impurity spectral function, with solid light-blue lines for the target spectrum of the original AIM, and dashed dark-blue lines for the result within the effective non-interacting model using the ACR. A small pole-broadening of width 0.01 is used for visualization in both cases. The auxiliary chain parameters are shown in the right panels (b,d). With $M=23$ auxiliary sites retained at half-filling and $M=37$ sites retained in the mixed-valence case, we see essentially perfect spectral reconstruction (the ACR spectrum is obtained from a physical model, so it is correctly normalized by construction, guaranteed causal, and in these cases the integrated error is less than 0.1\%). Because the spectrum is rather structured at small $N$, the auxiliary chain parameters also show quite some complexity. Interestingly, for larger systems as the target spectrum becomes smooth, the auxiliary chain becomes simpler.

In Fig.~\ref{fig:eq_mps} we show the same comparison but now for $N=101$ bath sites. The target AIM spectral function is obtained here using MPS methods. In practice we employ DMRG with TDVP~\cite{schollwock2006methods,Haegeman2016,Paeckel2019} within the ITensor package~\cite{fishman2022itensor}, using a DMRG ground state error threshold of $10^{-8}$ and a TDVP cutoff  of $3\cdot 10^{-7}$, with a timestep $\tau=0.2$ over a time range $t\in [0,1000]$. 
This time-dependent function is then Fourier transformed to yield the frequency-domain Green function. 
The Appendix presents a further analysis of the accuracy of the resulting MPS Green's function -- comparing with NRG in the $N\to \infty$ limit, and showing how the interaction self-energy can be directly obtained within the MPS approach. For the mapping, we use $M=140$ sites here with the ACR. The auxiliary chain parameters converge relatively rapidly with site index $n$. Longer auxiliary chains are required to reach very low energies, and the chain truncation therefore cuts off the narrow Kondo resonance, as seen in panel (a). However this is not a fundamental limitation but a practical consideration: arbitrary spectral structure can be resolved with long enough auxiliary chains, since any function can be expanded as a continued fraction in principle.

%%%%%%%%%%%%%%%%%%%%%%%%%%%
%%%%%%%%%%%%%%%%%%%%%%%%%%%

\section{Quench dynamics of the\\AIM occupation}\label{sec:quench}

We turn now to the non-equilibrium setting where the impurity potential of the AIM $\epsilon_d$ is quenched at $t=0$ from $\epsilon_d^i$ to $\epsilon_d^f$. For concreteness, we consider the specific case of quenching from half-filling $\epsilon_d^i=-U/2$ to mixed-valence $\epsilon_d^f=0$, and follow the time evolution of the impurity occupation $\langle n_d(t)\rangle$. Below we briefly describe how $\langle n_d(t)\rangle$ is computed in practice.

%%%%%%

\subsection{Exact diagonalization for short chains}
\label{ed_aim}
The impurity occupation $\langle n_d(t)\rangle$ of the AIM after a quench is straightforwardly computed using ED. This solution is of course only feasible for rather small system sizes. Here we use ED for $N=7$, as per Fig.~\ref{fig:eq_ed}.

First we define the real-space occupation basis $\{ |\phi_j\rangle \}$, where $|\phi_j\rangle = |n_{1\uparrow}^j, n_{1\downarrow}^j, n_{2\uparrow}^j, n_{2\downarrow}^j,  \ldots \rangle$ is the $j^{\text{th}}$ basis vector, such that $\hat{n}_d |\phi_j\rangle = n_d^j|\phi_j\rangle = \left( n_{d\uparrow}^j + n_{d\downarrow}^j\right) |\phi_j\rangle$. The time-independent Schr\"odinger equation is solved independently for the pre-quench $H^i$ and post-quench $H^f$ AIM Hamiltonians by diagonalization in this basis, $H^{i,f}|\psi^{i,f}_k \rangle=E^{i,f}_k|\psi^{i,f}_k \rangle$. The occupation basis states can be expressed in terms of these eigenstates $|\psi^{i,f}_k \rangle$ as 
$|\phi_j\rangle = \sum_k U^{i,f}_{jk} |\psi^{i,f}_k \rangle $.
The matrix $U$ is of dimension $4^D\times 4^D$ where $D$ is the total number of sites in the system (although in practice one exploits symmetries and quantum numbers to block-diagonalize the Hamiltonian).

The time-dependent, post-quench occupation of the impurity $d$ is then given by $\langle \ope n_d(t) \rangle = \text{tr}(\rho (t) \ope{n}_d) $, where the time-evolved density matrix is given by $\rho (t) = e^{-i t H^f } \rho_0\: e^{it H^f }$ with the equilibrium (pre-quench) density matrix $\rho_0 = e^{-\beta H^i}/Z_i$ and initial partition function $Z_i=\text{tr} ( e^{-\beta H^i})$ at inverse temperature $\beta=1/T$. The trace is evaluated using the eigenstates of $H^f$ to obtain,
\begin{eqnarray}\label{ntedaim}
	&\langle \ope n_d(t) \rangle = \qquad\qquad\qquad\qquad\qquad\qquad\qquad\qquad\qquad\qquad\\    &\dfrac{1}{Z_i}\dsum\limits_{jklp} n^j_d \, e^{i(E^f_l -  E^f_k )t} \, e^{-\beta E^i_p}
	U^{f}_{jl} U^{f}_{jk} 
	 \cdot \left( U^{f\dagger} U^i \right)_{lp} \left( U^{f\dagger}  U^i\right) _{kp} \nonumber
\end{eqnarray}

\begin{figure}
    \centering
    \includegraphics[width=0.75\linewidth]{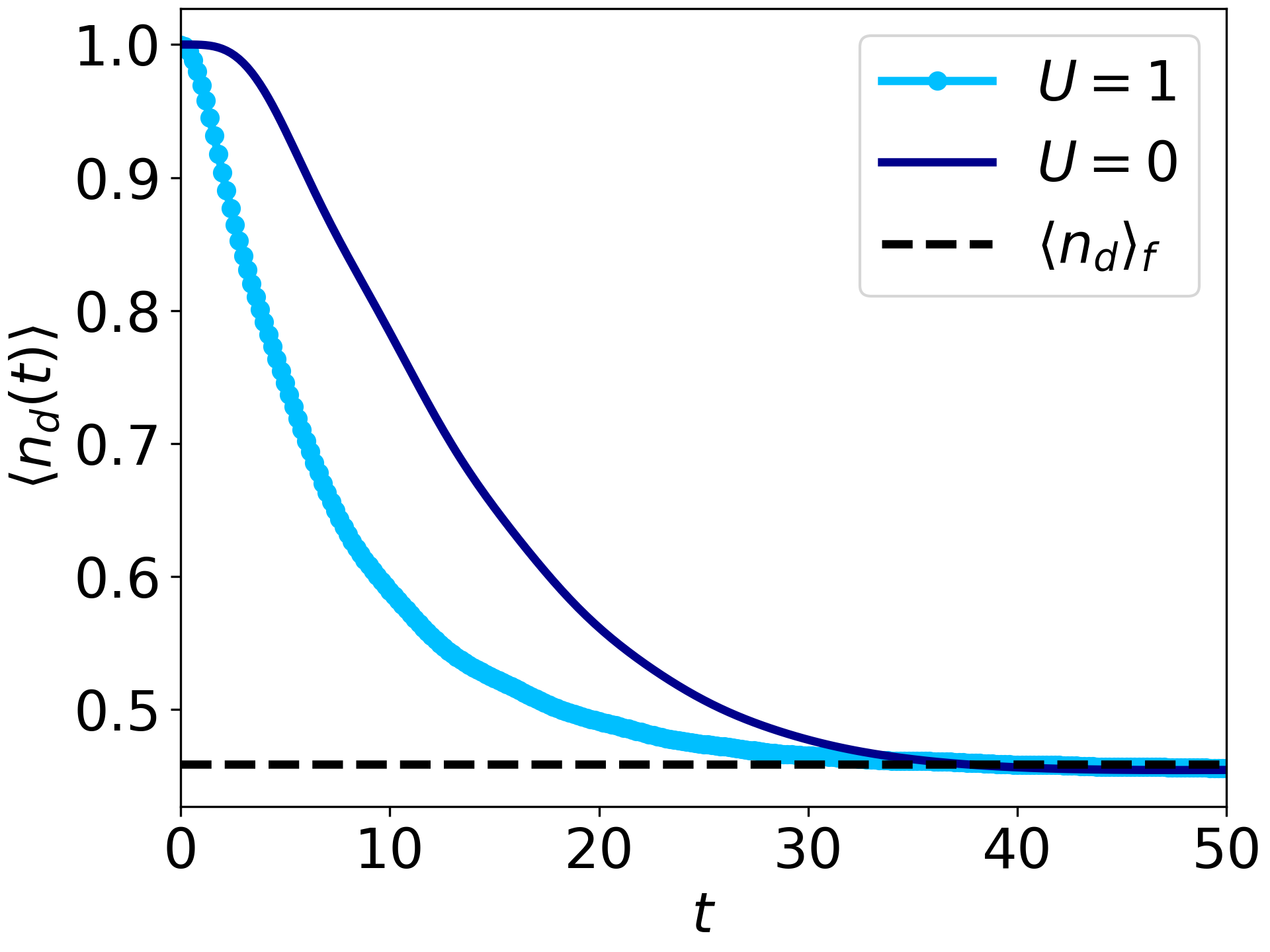}
    \caption{Time evolution of impurity occupation following a quench in the AIM, comparing interacting ($U=1$, light blue line) and non-interacting ($U=0$, dark blue) cases. Plotted for $N=301$, $t=0.5$, $V=0.2$, $\epsilon_d^i=-U/2$. For $U=1$ we choose $\epsilon_d^f=0$; for $U=0$ we tune $\epsilon_d^f$ to give the same impurity occupation in the long-time limit (dashed line).}
    \label{fig:AIMvsRLM}
\end{figure}

%%%%%%%%%%%

\subsection{MPS method}
In order to access the post-quench occupation dynamics for longer chain lengths, we employ MPS methods. At zero temperature we use DMRG to compute the ground state $|\psi_0^i \rangle$ of $H^i$, and then the time evolution under $H^f$ is generated using TDVP~\cite{schollwock2006methods,Haegeman2016,Paeckel2019}. The occupation number can then be readily evaluated as:
\begin{equation}
    \langle \ope n_d(t) \rangle= \ \langle \psi (t) |  n_d | \psi (t) \rangle
\end{equation}
where $| \psi (t) \rangle = e^{-it  H^f } |\psi_0^i \rangle$.

%%%%%%%%%%

\subsection{Formulation for Gaussian states}
For non-interacting Hamiltonians, such as the RLM or the effective model with the ACR, there are dramatic simplifications due to the underlying Gaussian structure~\cite{nghiem2018time}. 
The pre- and post-quench Hamiltonians $H^{i,f}$ corresponding to either Eq.~\ref{eq:AIMchain} at $U=0$ or Eq.~\ref{eq:Heff} can be diagonalized by canonical transformation of the fermionic operators, to give the form:
\begin{equation}
    H^{i,f}=\sum_k \epsilon^{i,f}_k \left(\psi^{i,f}_k\right)^\dagger \psi^{i,f}_k \;,
\end{equation}
where $\psi^{i,f}_k=\sum_l \tilde{U}^{i,f}_{lk} \lambda_l$ and $\lambda_l$ are the original fermionic operators $d_{\sigma}$, $c_{n\sigma}$ and $f_{n\sigma}$. We denote $l=d$ as the impurity site. In the chain forms of $H^{i,f}$ one obtains $\epsilon^{i,f}_k$ and $\tilde{U}^{i,f}_{lk}$ by diagonalizing the $D\times D$ tridiagonal adjacency matrix of coupling constants, where $D$ is the total number of sites in the system. This is equivalent to diagonalizing in the single-particle sector, and has an exponentially-reduced complexity relative to full diagonalization in the many-particle basis.

The time-dependence of the operators $\psi^f_k(t)$ after the quench can be evaluated using the Ehrenfest equation~\cite{ehrenfest1927bemerkung}
$\partial \psi_k^{f}(t) / \partial t = i [ H ^f,  \psi_k^{f}(t)]$. For Gaussian states, we can also write $\text{tr} (\rho_0 \psi_k^{i\dagger} \psi_{k'}^{i}) = \delta_{k k'} f(\epsilon_k^i)$, where $f(x)$ is the Fermi-Dirac distribution. The expression for $\langle n_d(t) \rangle$ therefore simplifies to,
\begin{eqnarray}\label{eq:ndt_nonint}
   & \langle n_d(t) \rangle = \qquad\qquad\qquad\qquad\qquad\qquad\qquad\qquad\qquad\qquad\\ & \sum_{knm} f(\epsilon_k^i) \ e^{i(\epsilon^f_n -\epsilon^f_{m} )t} \  \tilde U^{f}_{dn}\tilde U^{f}_{dm}  \cdot \left(\tilde U^{f\dagger} \tilde U^i \right)_{nk} \left(  \tilde U^{f\dagger} \tilde U^i\right)_{mk} \;. \nonumber
\end{eqnarray}
Evaluation of this expression is highly efficient.

%%%%%%%%%%

\begin{figure}
    \centering
    \includegraphics[width=1\linewidth]{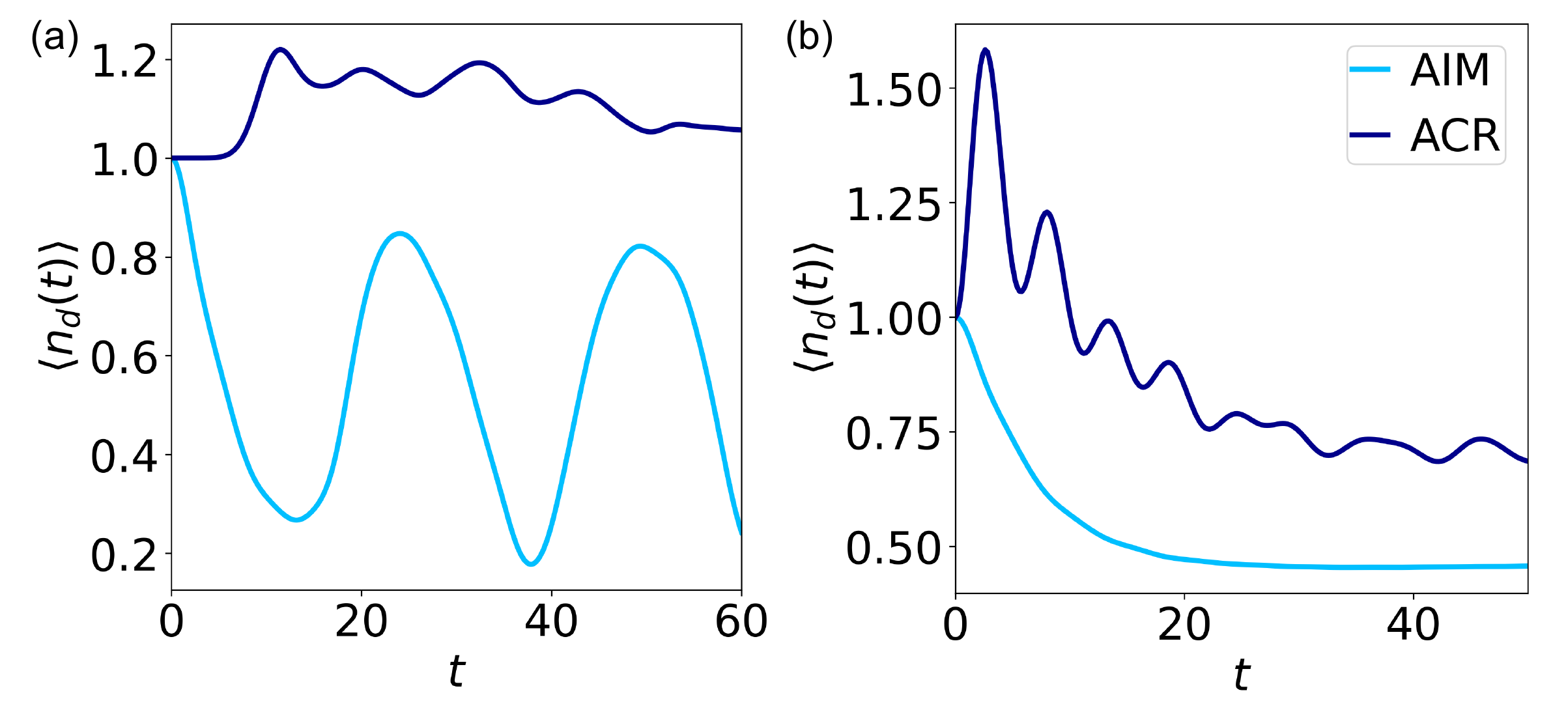}
    \caption{Dynamics of the impurity occupation in the AIM (light blue lines) for $N=7$ (a) and $N=101$ (b) bath sites, after quenching between the two configurations shown in Figs.~\ref{fig:eq_ed} and \ref{fig:eq_mps}. Dark blue lines show the  evolution obtained within the effective non-interacting model Eq.~\ref{eq:Heff}, by quenching between pre- and post-quench \textit{equilibrium} auxiliary chains.}
    \label{fig:eq_aux_quench}
\end{figure}

\subsection{Equilibration and the role of interactions}
As an illustration, consider a potential quench in the interacting AIM with long chain length $N=301$ sites at zero temperature. The resulting time-dependence of the impurity occupation is shown as the light blue line in Fig.~\ref{fig:AIMvsRLM}. The horizontal dashed line shows the equilibrium occupation of $H^f$, to which $\langle n_d(t)\rangle$ tends at long times. Over the timescales considered, this shows effective equilibration of the system (although eventually recurrences appear in any finite-size system simulation). For comparison, consider the same system but with interactions switched off, $U=0$. Here we choose $\epsilon_d^f$ specifically so that $\langle n_d\rangle_f$ matches the interacting $U=1$ calculation. As shown by the dark blue line, the relaxation is much slower, with pronounced differences in the evolution just after the quench. This is because interactions produce additional scattering channels that allow for faster equilibration. This raises the question as to whether the interacting quench dynamics can be captured by a purely non-interacting system with additional scattering channels. This is the philosophy with the auxiliary chain representation, where in equilibrium, the interaction self-energy was replaced by a secondary bath of non-interacting fermions. Does this straightforwardly generalize out of equilibrium?

%%%%%%%%%%

\subsection{Quench between equilibrium auxiliary chains}
A naive first guess to capture the non-equilibrium dynamics due to a quench from interacting Hamiltonians $H^i$ to $H^f$ within a purely non-interacting model might be to simply quench between the two \textit{equilibrium} auxiliary chain forms. 

Consider a quench in the interacting AIM Eq.~\ref{eq:AIMchain} with $N=7$ from $\epsilon_d^i=-U/2$ to $\epsilon_d^f=0$. The equilibrium impurity spectral functions corresponding to $H^i$ and $H^f$ were shown in Fig.~\ref{fig:eq_ed}(a,c). The resulting time evolution of the impurity occupation is shown as the light blue line in  Fig.~\ref{fig:eq_aux_quench}(a), computed directly with ED. It shows strong oscillations due to severe finite-size effects of the short chain. If we instead take the non-interacting effective model in the ACR form Eq.~\ref{eq:Heff}, and quench between the equilibrium auxiliary chain parameters shown in Fig.~\ref{fig:eq_ed}(c,d), then we get the dark blue line in Fig.~\ref{fig:eq_aux_quench}(a). We can likewise repeat this for longer chains, for example $N=101$ whose equilibrium mapping was depicted in Fig.~\ref{fig:eq_mps}. The comparison of quench dynamics in the AIM and ACR in this case is shown in Fig.~\ref{fig:eq_aux_quench}(b).

Clearly in both cases the time-evolution resulting from quenching between \textit{equilibrium} auxiliary chains for $H^i$ and $H^f$ fails to capture the true non-equilibrium dynamics of the AIM. Does a non-interacting auxiliary chain representation of $H^f$ exist out of equilibrium? We show in the next section that by optimizing the auxiliary chain parameters of the post-quench effective model, the behavior of $\langle n_d(t)\rangle$ in the AIM can be exactly reproduced in a purely non-interacting setup.

%%%%%%%%%%%%%%%%%%%%%%%
%%%%%%%%%%%%%%%%%%%%%%%

\begin{figure}
    \centering
    \includegraphics[width=1\linewidth]{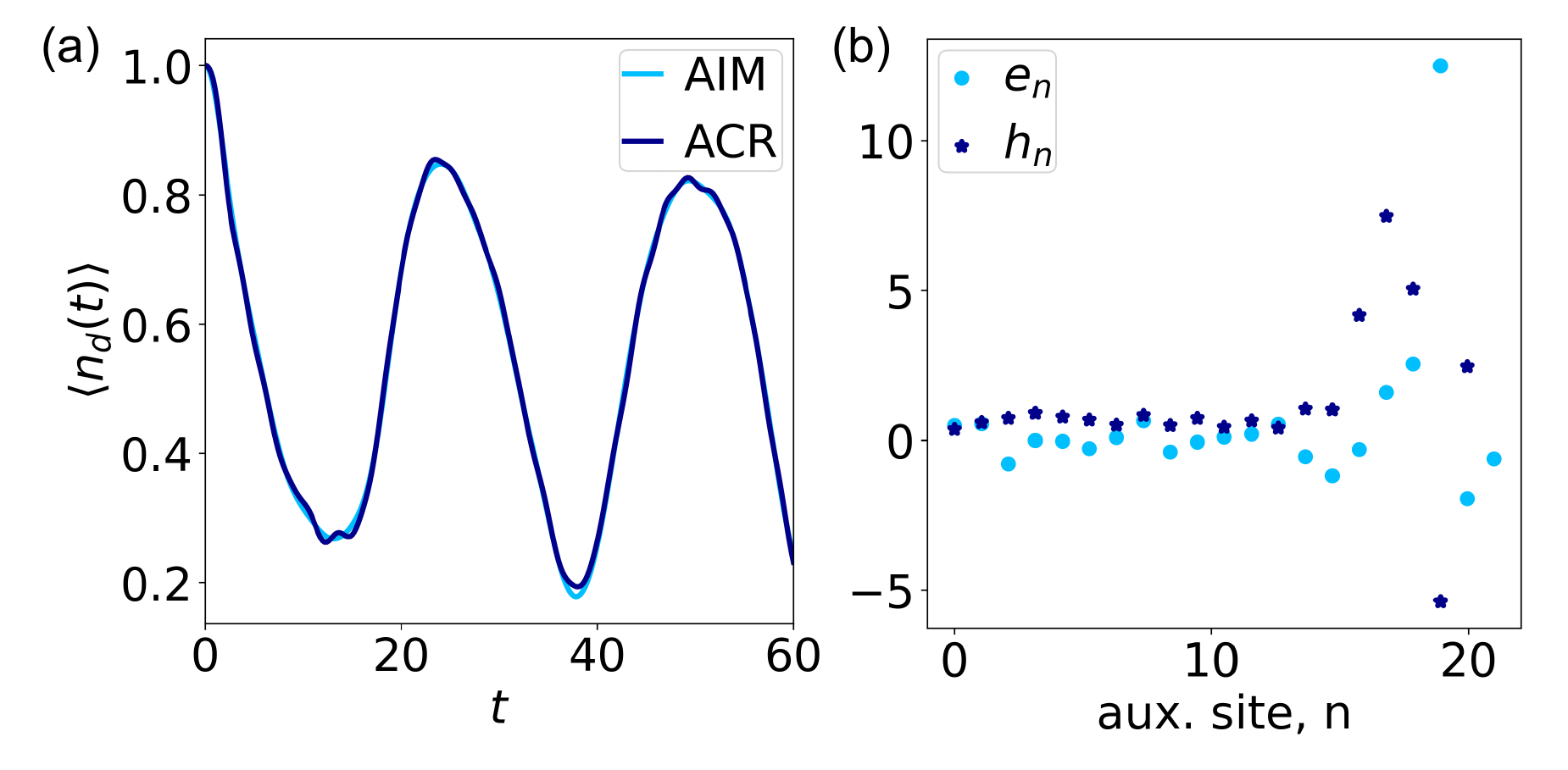}
    \caption{Quench dynamics of the impurity occupation in the interacting AIM with $N=7$ (light blue line) compared with results from the non-interacting ACR (dark blue) shown in panel (a), using optimized post-quench auxiliary chain parameters (panel b). $U=1$, $t=0.5$, $V=0.2$, $\epsilon_d^i=-U/2$, $\epsilon_d^f=0$.}
    \label{fig:MLopt_N7}
\end{figure}

\section{Auxiliary chain representation for the quenched AIM}

\begin{figure}[t]
    \centering
    \includegraphics[width=1\linewidth]{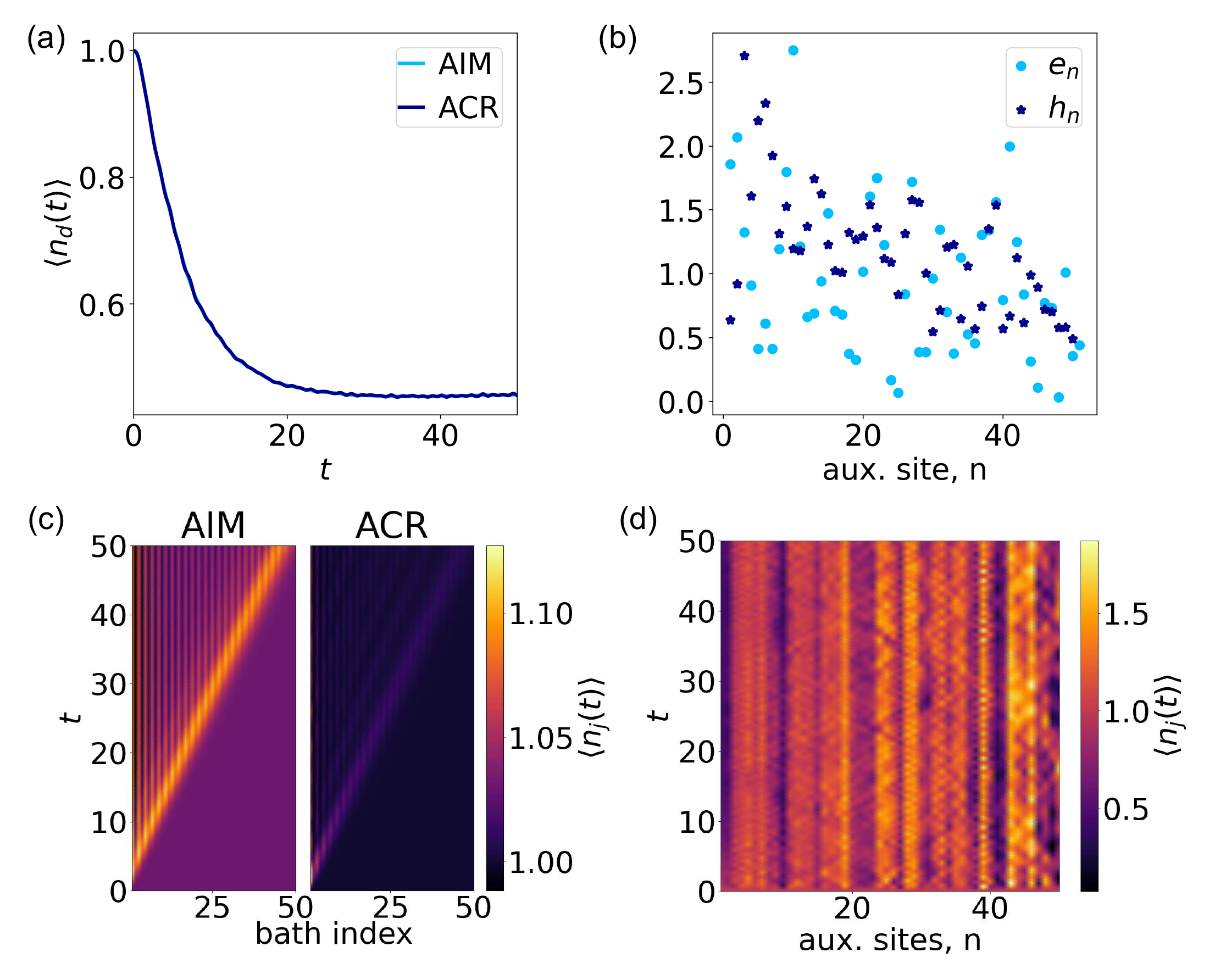}
    \caption{(a) Time-dependence of post-quench impurity occupation in the AIM (light blue) for the same AIM parameters as in Fig.~\ref{fig:MLopt_N7} but with $N=101$ sites. Results for the ACR (dark blue) obtained by optimizing the auxiliary chain parameters (shown in panel b) trained on impurity data only. (c) Evolution of occupations along the physical bath chain, comparing AIM and ACR. (d) Dynamics of auxiliary chain occupations.}
    \label{fig:MLopt_imp}
\end{figure}

We seek to find an  ACR for the post-quench Hamiltonian $H^f$ such that the true behavior of $\langle n_d(t)\rangle$ for the quenched AIM is exactly reproduced. Since the impurity Green's function and occupation for $t<0$ before the quench is given by the equilibrium ACR of $H^i$, we take this to be our pre-quench effective model. Without any obvious deterministic way of finding the auxiliary chain parameters of $H^f$ out of equilibrium, we resort to numerical optimization methods. We treat the effective ACR parameters $\tilde{V}$, $\tilde{\epsilon}_d$, $\{e_n\}$ and $\{h_n\}$ (collectively denoted by $\{\xi_i\}$) as variational parameters to be numerically optimized by minimization of some objective loss function. To capture the time-dependence of the post-quench impurity occupation, we choose the loss function,
\begin{equation}\label{eq:loss}
    \mathcal{L}=\frac{1}{R}\sum_{x=1}^{R}\left ( \langle n_d(t_x)\rangle^{AIM}-\langle n_d(t_x)\rangle^{ACR}\right )^2 + \frac{\Gamma^2}{P}\sum_{i=1}^P \xi_i^2 
\end{equation}
where $t_x$ is a discrete time step in the range $[0,t_{max}]$ with step size $\delta t=0.2$. The second term punishes large values of effective model parameters, and prevents runaway flow of the optimization to unphysical solutions. In practice we choose $\Gamma=10^{-3}$. The loss function is minimized by optimizing the effective model parameters $\xi_i$ using Powell's (local) conjugate direction method~\cite{powell1964efficient, 2020SciPy-NMeth}. 
We run the algorithm for 1000 iterations before applying a small perturbation to the parameters and then restarting the optimization. This helps to avoid convergence in local minima. We typically restart the algorithm in this way 30--40 times and take the optimized solution as the one with lowest loss across all runs.

As an illustration, consider Fig.~\ref{fig:MLopt_N7} for an AIM with $N=7$. The optimization yielded a mean-square-error (MSE) of less than $10^{-4}$, resulting in the auxiliary chain parameters shown in panel (b). The time-evolution of the impurity occupation $\langle n_d(t) \rangle$ computed from the effective non-interacting model matches accurately the true behavior of the interacting AIM, see panel (a). The evolution was captured at this level using $M=20$
auxiliary chain sites. The optimized auxiliary chain parameters for the first $~15$ sites take modest values, before becoming comparatively large towards the end of the auxiliary chain. These large values are apparently required to reach convergence below the loss function threshold of $10^{-4}$ for this setup.

The auxiliary chain form of our non-interacting effective model naturally organizes information about the post-quench dynamics hierarchically, exploiting light-cone effects that constrain information spreading. Short-time behavior is mapped to auxiliary sites near the impurity, whereas long-time behavior is mapped to the more distant sites. Since the entire auxiliary chain is quenched simultaneously at $t=0$, information about the change in the parameters at auxiliary site $n$ takes a time $t\sim n/\bar{h}$ to propagate back to the impurity~\cite{Lieb1972,Bravyi2006}, where $\bar{h}$ is the typical chain hopping. Therefore the time-series data for $\langle n_d(t) \rangle$ can roughly be thought of as being encoded in the spatial evolution of the auxiliary chain parameters.

\begin{figure}[t]
    \centering
    \includegraphics[width=1\linewidth]{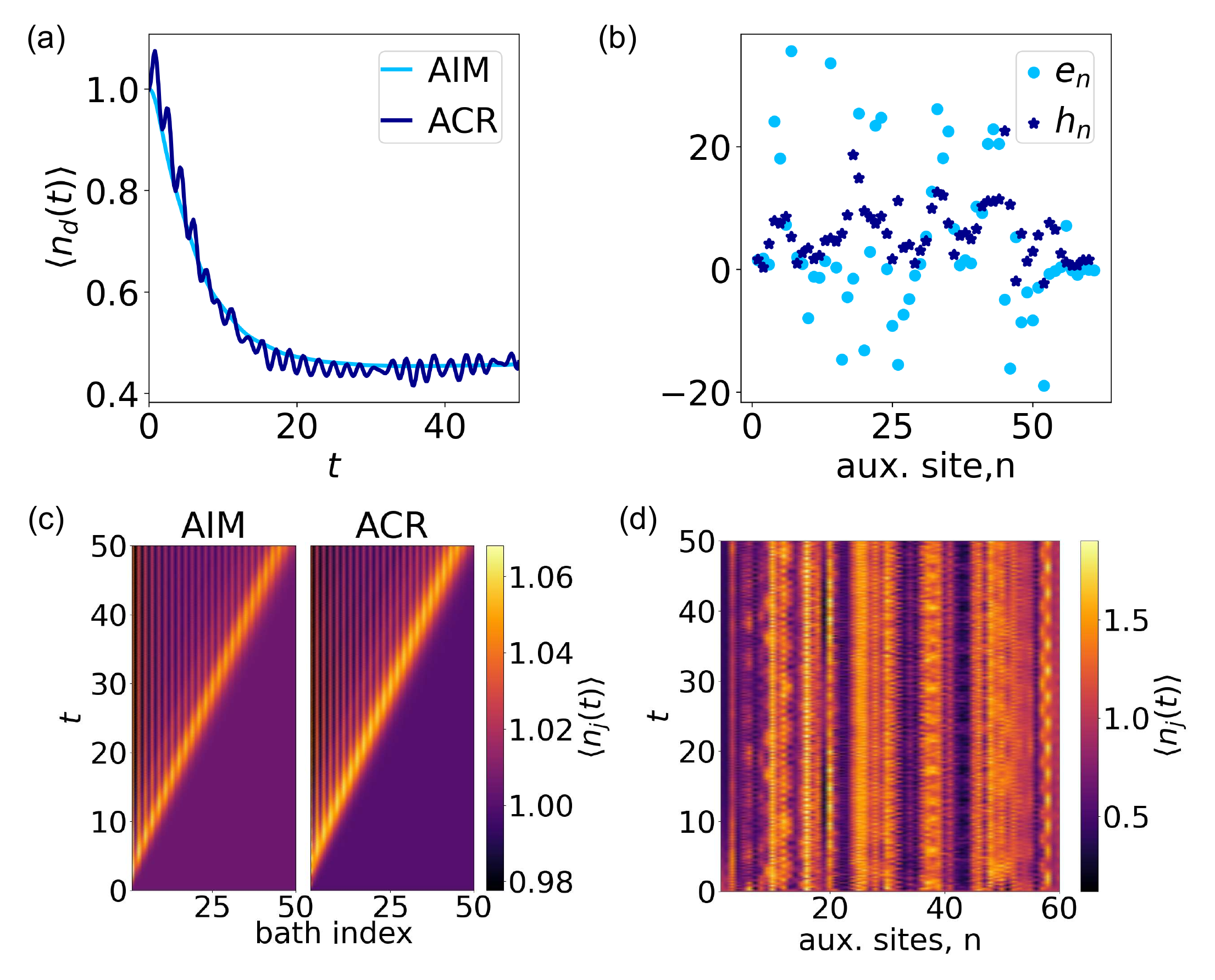}
    \caption{Same as Fig.~\ref{fig:MLopt_imp} but now the ACR is trained on post-quench AIM occupation data for all physical sites.}
    \label{fig:MLopt_all}
\end{figure}

Fig.~\ref{fig:MLopt_imp}(a,b) shows results for an AIM with $N=101$ sites. The bare model was solved within TDVP-MPS as before to obtain $\langle n_d(t) \rangle$, which was the input to train the auxiliary chain parameters of the effective model. Here we use an auxiliary chain of length $M=50$ and convergence threshold on the MSE of $10^{-6}$. 
We start the optimization from an initially flat parameter configuration $\{\xi_i\}^i=1$, which helps avoid 
local minima. 
The optimized parameters of the ACR are shown in panel (b), and the resulting time-evolving impurity occupation is compared against the target AIM result in panel (a). The ACR is shown to be able to capture the true impurity quench dynamics very accurately.

Within the physical system of the target AIM system, one effect of the potential quench on the impurity at time $t=0$ is to send an electronic occupancy ``ripple'' which propagates along the bath chain, away from the impurity, for $t>0$. The resulting light-cone effect~\cite{Lieb1972,Bravyi2006} is seen very clearly in the left panel of Fig.~\ref{fig:MLopt_imp}(c), which depicts the time-dependent occupancy $\langle n_j(t)\rangle$ of bath site $j\in [1,N]$. The right panel of Fig.~\ref{fig:MLopt_imp}(c) shows that this behavior is \textit{not} correctly captured by the ACR effective model. Training the effective model on data for the impurity alone does not guarantee the correct quench dynamics for the full physical system. 

The time-dependent occupations $\langle n_j(t)\rangle$ of sites $j$ in the \textit{auxiliary} chain are shown for comparison in Fig.~\ref{fig:MLopt_imp}(d). Interestingly, we see relatively modest temporal fluctuations, but rather strong spatial fluctuations in the site occupancy following the quench. We will return to the physical interpretation of the ACR structure  shortly.

Together, Fig.~\ref{fig:MLopt_imp} shows that an ACR for a non-interacting effective model can be found that matches the impurity occupation, but not the full (single-particle) quench dynamics of the physical system (impurity plus original bath sites). The optimization of the set of parameters $\{\xi_i \}$ is in this sense underconstrained by the single function $\langle n_d(t)\rangle$. 

\begin{figure}[t]
    \centering
    \includegraphics[width=1\linewidth]{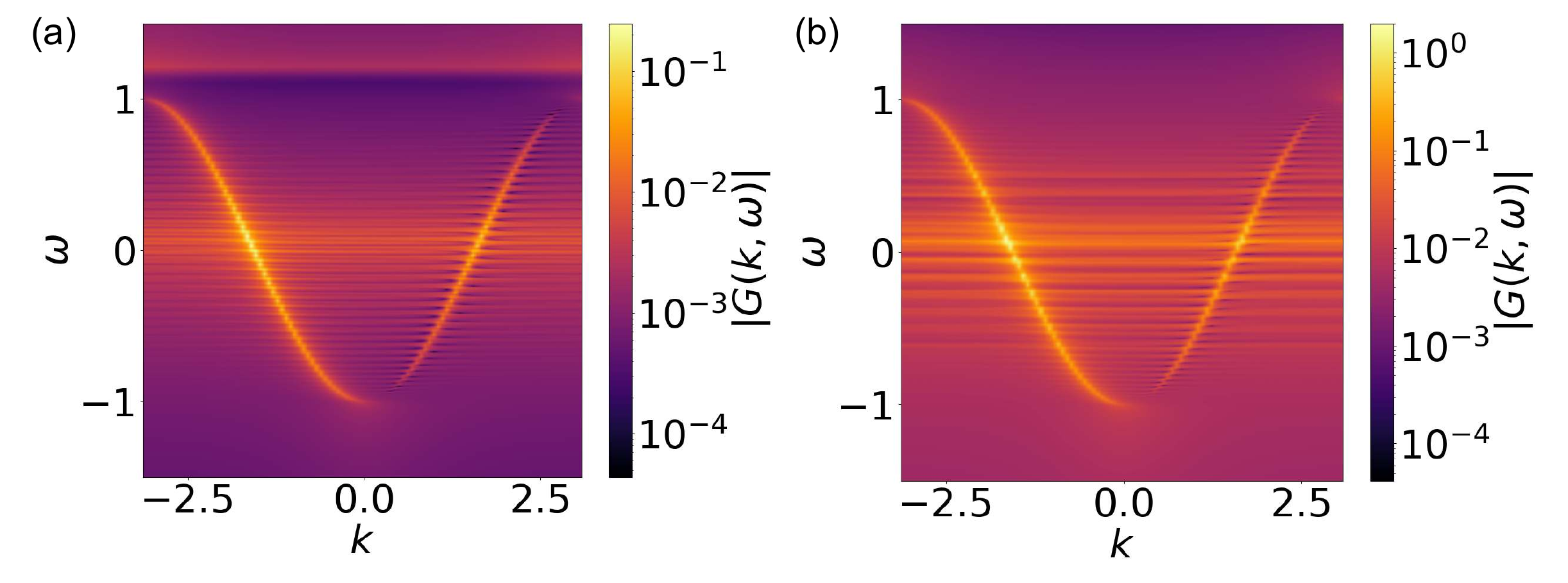}
    \caption{Band structure of the interacting AIM (a) and optimized non-interacting ACR (b) corresponding to Fig.~\ref{fig:MLopt_all}. Real-space Green's functions $G_{
    dn;\sigma}(\omega)$ on physical bath sites $n\in [1,N]$ with $N=101$ are Fourier transformed to momentum space $G(k,\omega)$ and plotted as a colormap vs $\omega$ and $k$.}
    \label{fig:disp}
\end{figure}

A more ambitious approach to training the effective model involves optimizing the ACR parameters $\{\xi _i \}$ so as to minimize the MSE for the time-dependent occupations on \textit{all sites} (impurity plus bath), generalizing Eq.~\ref{eq:loss}. Results of this strategy are shown in Fig.~\ref{fig:MLopt_all}, when keeping $M=60$ auxiliary sites and using the training protocol described previously. The lowest MSE obtained in this case was $\sim 10^{-5}$, where we ran 30 Powell iterations of 1000 runs per iteration. 
The time-dependence of the impurity occupation $\langle n_d(t)\rangle$ in Fig.~\ref{fig:MLopt_all}(a) is somewhat noisy, with small spurious high-frequency oscillations around the true evolution. However, the basic features are correctly reproduced by the trained ACR parameters shown in panel (b). Importantly, as shown in panel (c), the correct evolution of the real-space occupations is also captured, showing the coherent light-cone effect. The auxiliary chain site occupations in panel (d) again show strong spatial but modest temporal fluctuations. 
The good agreement between AIM and effective models, with the latter trained on the evolution of all physical occupations, indicates that the ACR is indeed expressive enough to describe the quench dynamics of this system. We anticipate that the auxiliary chain length and training protocol could be optimized to reduce complexity of the effective model parameters and increase fidelity with the target dynamics. We leave further development to future work. 

Although the effective model training was done on the level of the site occupation expectation values $\langle n_j(t)\rangle$, one can ask whether the same optimized effective model captures other physical properties. In Fig.~\ref{fig:disp}(a), we consider the Fourier transform of the site-dependent retarded Green's functions $G^{\rm AIM}_{dn;\sigma}(\omega)$ of the target AIM, where $n$ is a bath index $n\in [1,N]$ and again $N=101$. The full frequency dependence of the real-space Green's functions is computed in the \textit{equilibrium} post-quench AIM model $H^f$ using TDVP-MPS, and then Fourier transformed to momentum space to obtain $G(k,\omega)$. This is plotted as a color map in Fig.~\ref{fig:disp}(a). Two physical features are of note. Firstly, we see clearly the 1d dispersive cosine band structure of the bath chain. Secondly, we see an enhancement of spectral weight around the Fermi energy $\omega=0$ for all momenta, corresponding to the Kondo effect (cf.\ Fig.~\ref{fig:eq_mps}(c) for the impurity site spectrum). In Fig.~\ref{fig:disp}(b) we show the corresponding result obtained in the long-time limit from the all-sites-optimized effective ACR model. Despite some spurious frequency-dependent oscillations, the ACR model is seen to capture the key dynamical structures. This is nontrivial because the effective model was trained only on one-point expectation values, not two-point dynamical correlation functions. This adds further weight to the conclusion that the ACR can be an expressive and accurate effective model for the AIM quench dynamics -- at least on the level of single-particle dynamical quantities.

%%%%%%%%%%%%%%%%%%%%%%%%%%%%%
%%%%%%%%%%%%%%%%%%%%%%%%%%%%%

\begin{figure}[t]
    \centering
    \includegraphics[width=1\linewidth]{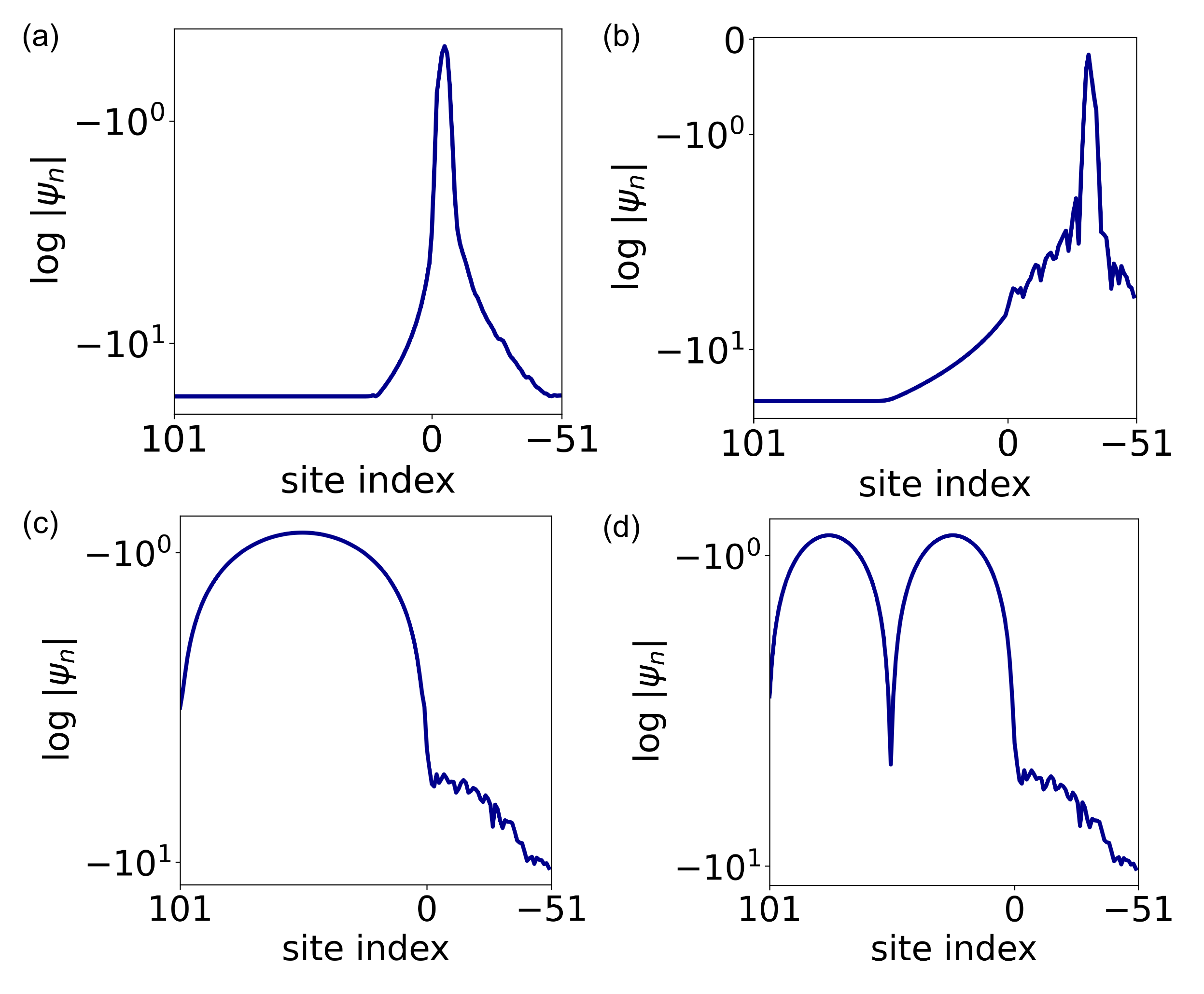}
    \caption{Single-particle wavefunction amplitudes for the optimized ACR corresponding to Fig.~\ref{fig:MLopt_all}. Plotted are the ground state in (a) and the $8^{th}$, $9^{th}$ and $10^{th}$ excited states in (b,c,d). Here the impurity is labelled as site index zero, while physical bath sites are $n\in [1,N]$ and auxiliary chain sites are $n\in [-1,-M]$ with $N=101$ and $M=60$ as before.}
    \label{fig:evects}
\end{figure}

\section{Discussion and conclusions}\label{sec:conc}
We have argued that the auxiliary chain mapping from the interacting AIM to an effective Gaussian theory (which can be rigorously derived in equilibrium) can be generalized to out-of-equilibrium settings. We demonstrate by direct calculation that the real-time quench dynamics of physical site occupations in the AIM can be captured by a non-interacting version of the same model, but with additional coupling to an auxiliary chain of non-interacting fermions. A potential quench in the original AIM then corresponds to quenching the parameters of this auxiliary chain.

A related strand of work also represents correlated impurities using auxiliary non-interacting sites. In the auxiliary master equation approach~\cite{Arrigoni2013,Dorda2014} and Lindblad-driven / pseudomode schemes~\cite{Dorda2017,Thoenniss2024}, the continuum reservoir is replaced by a few auxiliary bath sites coupled to a Markovian environment, fitted to the hybridization function. There the interaction is retained and the auxiliary sites encode the bath. Here, by contrast, the interaction is removed entirely and the auxiliary chain encodes the self-energy, yielding a strictly Gaussian — and closed, rather than dissipative — effective model. 

Another related approach is the ghost Gutzwiller approximation~\cite{Frank2024,MejutoZaera2024,Giuli2026}, where auxiliary `ghost' orbitals encode a finite-pole self-energy; its time-dependent extension~\cite{Guerci2023} can be applied to interaction quenches of the Hubbard model. In that approach the embedding Hamiltonian contains the impurity interaction and is solved as an interacting (but small) many-body problem.

Within our auxiliary chain approach, the real-time dynamics of the effective non-interacting system is extremely efficiently computed, compared with that of the original interacting AIM, which requires sophisticated many-body computational techniques. However, we do not claim that the existence of this mapping provides a shortcut route to solving out-of-equilibrium many-body problems, since the mapping itself must be deduced from dynamical data obtained from solution of the original model. Rather, the existence of the mapping provides a novel lens through which to view the complicated non-equilibrium dynamics of quenched interacting systems. In particular, the structure and geometry of the auxiliary chain setup is designed to map the continuous real-time behavior of the system after the quench to the spatial evolution of a finite set of coupling parameters along the chain, due to light-cone effects. Indeed, our auxiliary chain model can itself be regarded as a physically-motivated and interpretable machine-learning model, which can be efficiently optimized to generate the correct functional form of the AIM occupation dynamics. That is, instead of using, say, a classical Boltzmann machine with tuned weights, one can use the auxiliary chain model with optimized chain parameters. Because the ACR is Gaussian, each function evaluation is intrinsically low-cost. It would be an interesting next step to investigate whether the ACR parameters themselves could be generatively predicted using a neural network trained on a dataset of different AIM quenches.

A potential advantage of the ACR to model quench dynamics is that it affords the opportunity to understand and interpret the optimized solution physically, in a way that is typically not possible for standard `black box' machine leaning methods. In our setup, the physical system (with interactions switched off) is coupled at the impurity to a quenched auxiliary system. The auxiliary system has its own nontrivial real-time dynamics following the quench. From the perspective of the impurity, one can view this as providing a specific \textit{time-dependent modified boundary condition}. 

In this respect, one striking feature of the optimized ACR parameters -- see e.g.\ Fig.~\ref{fig:MLopt_all}(b) -- is that they appear rather random or chaotic; yet this structure seems necessary to recover the correct system evolution. The strong spatial modulations of the auxiliary chain occupations shown in Fig.~\ref{fig:MLopt_all}(c) are suggestive of localization in the auxiliary chain. This picture is supported by Fig.~\ref{fig:evects}, where we plot the wavefunction amplitude for representative low-energy states in the non-interacting effective model. Here we use a site-index notation in which the impurity is labelled as site zero, the physical bath sites are $n\in [1,N]$ with $N=101$, and the auxiliary chain sites are $n \in [-1,-M]$ with $M=60$ as per Fig.~\ref{fig:MLopt_all}. We see that wavefunctions have support largely on \textit{either} the physical system sites (lower panels) or on the auxiliary chain sites (upper panels), with only exponentially-small leakage between the two subsystems. The wavefunctions with support on system sites in the lower panels are extended and are typical of metallic 1d chain solutions. By contrast, wavefunctions with support on the auxiliary chain appear strongly localized, so that physical and auxiliary sites effectively decouple at low energies.

This decoupling, and the localized character of the auxiliary states more generally, may arise from more than one mechanism. The comparatively large values of the post-quench auxiliary chain parameters generate both an apparent disorder along the chain, pointing towards Anderson localization, and a substantial mismatch in onsite potential between the physical bath and the auxiliary chain across their shared boundary, which could itself induce localized states near the interface. Together these offer an appealing picture of how information injected at the impurity propagates into the auxiliary chain, is partly trapped there, and is gradually released back into the physical bulk -- effectively controlling the flow of information between the auxiliary system and the physical sites. Disentangling the relative roles of disorder and boundary-potential mismatch would be an interesting direction for further study.

Aside from the proof-of-principle demonstrations presented here, it would be interesting to systematically investigate the minimum auxiliary chain complexity required to converge solutions for different quenches, and improved optimization protocols that exploit the light-cone physics of the problem. Different quench problems could be studied, and different impurity models considered. The mapping could be investigated for the case of finite-time driving. It may be interesting to consider quenched lattice models, such as the Hubbard model, within the context of DMFT. We leave these open directions for future work.

\begin{figure}[t]
    \centering
    \includegraphics[width=1\linewidth]{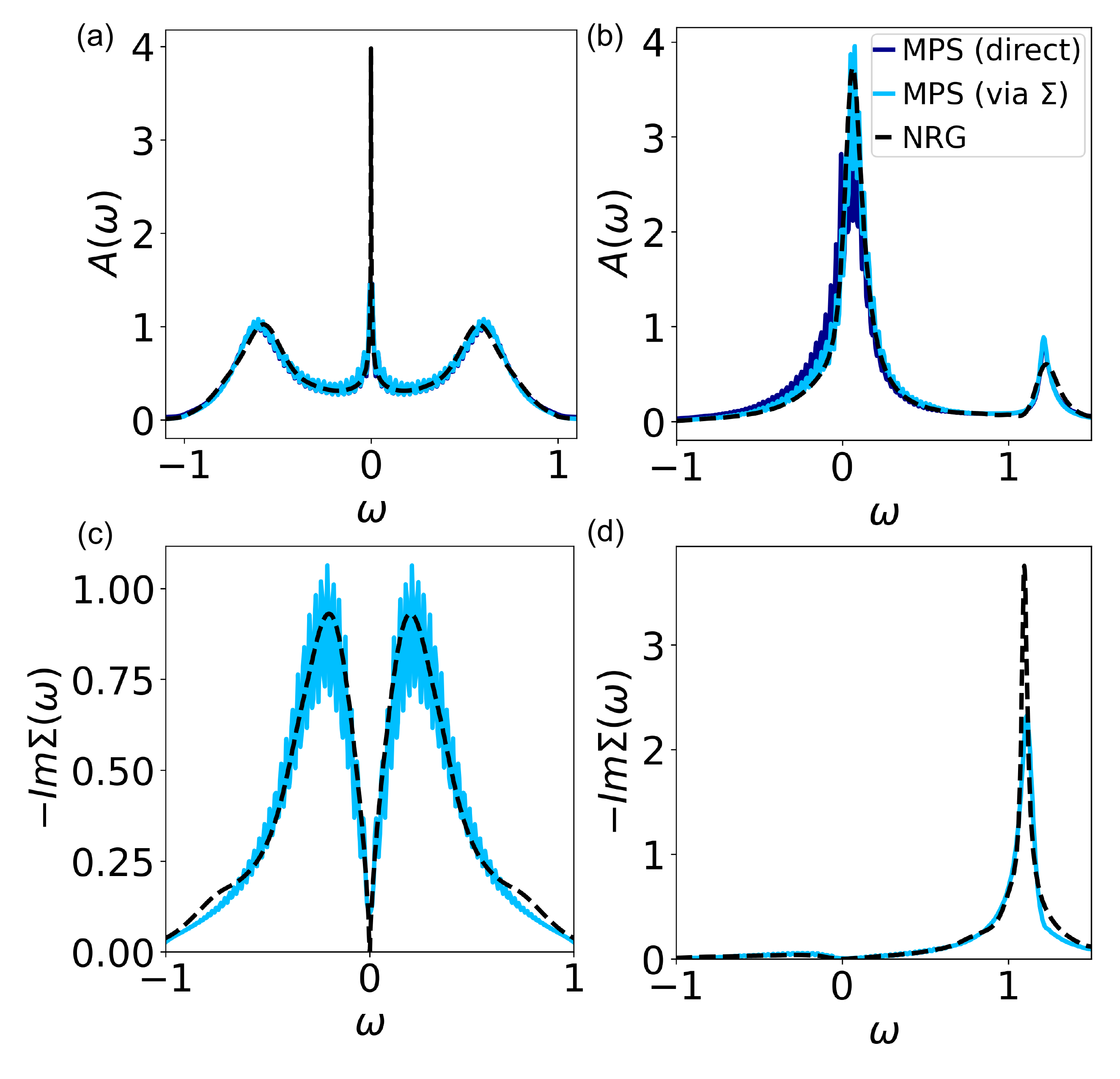}
    \caption{Benchmark comparison of the equilibrium impurity spectral function $A(\omega)$ vs $\omega$ at $T=0$ (a,b) obtained within TDVP-MPS for $N=101$ bath sites (solid lines), compared with NRG results in the thermodynamic limit $N\to \infty$ (dashed lines). Plotted for $U=1$, $t=0.5$, $V=0.2$ and $\epsilon_d=-U/2$ (a) or $\epsilon_d=0$ (b). 
    Panels (c,d) show the corresponding self-energies obtained from Eq.~\ref{eq:se}. Dark-blue lines in (a,b) correspond to the direct evaluation of $A(\omega)$ using TDVP-MPS, whereas light blue lines are obtained from the Dyson equation Eq.~\ref{eq:dyson} via the self-energy.}
    \label{fig:app_GF}
\end{figure}

%%%%%%%%%%%%%%%%%%%%%%%%%%
%%%%%%%%%%%%%%%%%%%%%%%%%%

\begin{acknowledgments}
We acknowledge fruitful discussions with Ciar\'an Hickey.
EB acknowledges funding from the Irish Research Council Grant EPSPG/2022/59 partnered with Intel Research \& Development Ireland Limited, and the University College Dublin School of Physics Preston Scholarship. AKM acknowledges funding from Research Ireland through grant 24/FFP-P/12816.
\end{acknowledgments}

%%%%%%%%%%%%%%%%%%%%%%%%%%
%%%%%%%%%%%%%%%%%%%%%%%%%%

\appendix

\section{Self-energy within TDVP-MPS}
\label{SEappendix}

The equilibrium interaction self-energy $\Sigma^{\rm AIM}_{dd;\sigma}(z)$ of the AIM is defined by the Dyson equation, Eq.~\ref{eq:dyson}. Computing the impurity Green's function $G^{\rm AIM}_{dd;\sigma}(z)=\langle\langle d_{\sigma};d_{\sigma}^{\dagger}\rangle\rangle$ directly within TDVP-MPS~\cite{schollwock2006methods,Haegeman2016,Paeckel2019,fishman2022itensor} and then inverting Eq.~\ref{eq:dyson} to obtain the self-energy typically yields somewhat inaccurate results. Instead, the self-energy can be computed directly, following the prescription in Ref.~\cite{bulla1998numerical},
\begin{equation}\label{eq:se}
    \Sigma^{\rm AIM}_{dd;\sigma}(z)=\frac{U F^{\rm AIM}_{dd;\sigma}(z)}{G^{\rm AIM}_{dd;\sigma}(z)} \;,
\end{equation}
where $F^{\rm AIM}_{dd;\sigma}(z)=\langle\langle d_{\sigma}n_{d\bar{\sigma}};d_{\sigma}^{\dagger}\rangle\rangle$. We find that computing the Green's function via the self-energy from Eq.~\ref{eq:se} gives improved accuracy at low frequencies. A benchmark against state-of-the-art NRG calculations for the AIM in the thermodynamic limit $N\to \infty$ is presented in Fig.~\ref{fig:app_GF}. In the present work, we use the TDVP-MPS self-energy to compute the auxiliary chain parameters via its continued-fraction expansion.

%%%%%%%%%%%%%%%%%%%%%%%%%%
%%%%%%%%%%%%%%%%%%%%%%%%%%

\normalem % removed non-wrapping underlined text
%\bibliography{quench-paper.bib}

\providecommand{\noopsort}[1]{}\providecommand{\singleletter}[1]{#1}%

\end{document}